\documentstyle[12pt]{article}
\begin{document}
\title{\vspace*{-1.8cm}
Zweig-rule-satisfying inelastic rescattering in $B$ decays to
pseudoscalar mesons}
\author{
{P. \L{}ach and P. \.Zenczykowski}$^*$\\
\\
{\em Dept. of Theoretical Physics},
{\em Institute of Nuclear Physics}\\
{\em Radzikowskiego 152,
31-342 Krak\'ow, Poland}\\
}
\maketitle
\begin{abstract}
We discuss all contributions from Zweig-rule-satisfying SU(3)-symmetric 
inelastic 
FSI-induced corrections in $B$ decays to $\pi \pi$, $\pi K$, $K\bar{K}$,
$\pi \eta (\eta ')$, and $K \eta (\eta ')$ . It is shown how all of these
FSI corrections
lead to a simple redefinition of the amplitudes, permitting the use of 
a simple
diagram-based description, in which, however, 
weak phases may enter in a modified way. 
The inclusion of FSI corrections admitted by the present data 
allows an arbitrary
relative phase between the penguin and tree short-distance
amplitudes.
The FSI-induced error of the method, in which the value of
the weak phase $\gamma $ is to be determined by combining future results from 
$B^+,B^0_d,B^0_s$ decays to $K\pi$, is estimated to be of the order of 
$5^o$ for $\gamma \approx 50^o-60^o$. 
\end{abstract}
$^*$ E-mail:
zenczyko@iblis.ifj.edu.pl
\newpage

\section{Introduction}
Most of the analyses of CP-violating effects in $B$ decays deal
with quark-diagram short-distance (SD) amplitudes and assume that final
state interactions (FSI) are negligible.
On the other hand, it has been argued that this neglect is not justified, and
that any reliable analysis of $B \to PP$ decays
($P$ - pseudoscalar mesons)
  must take rescattering into account
\cite{Wolf,FSI}. While only small effects of rescattering through
low-lying 
intermediate states are generally expected, 
 the sequence 
$B \stackrel{weak}{\to} i \stackrel{FSI}{\to} PP$ with {\em inelastic}
multiparticle intermediate states $i$ might in principle 
lead to important corrections
\cite{Wolf}.
The estimates of the size of such inelastic effects are model-dependent,
 with the contributions from
different intermediate states either cancelling in an approximate way, or
having random phases \cite{Wolf}, or even adding coherently \cite{Zen2001}.

With our insufficient knowledge of $PP$ interactions at $5.2~GeV$, there is
virtually no hope that the rescattering effects may be reliably calculated.
Under the circumstances it seems appropriate to use
symmetry-based approaches and to
parametrize the rescattering in terms of a few FSI-related
parameters.
Such an approach has been recently studied in ref.\cite{Zen1}, where
SU(3) symmetry was used to reduce our ignorance of the
inelastic rescattering to a small set of effective
SU(3) parameters jointly describing all $i \to PP$ processes.
In the present paper we limit the considerations of ref.\cite{Zen1}
to the case when the rescattering amplitudes satisfy Zweig rule.

\section{Short-distance amplitudes}
Short-distance decay amplitudes lead to $q\bar{q} q\bar{q}$ states
which convert to hadron-level two-body states composed
of various mesons $M_1$ and $M_2$ (including the pseudoscalar mesons
$P_1$ and $P_2$).
Transitions to many-body states occur when the $q\bar{q}$  
states radiate off further quark-antiquark pairs and gluons
leading to the decays of $M_i$. In order to estimate the rescattering through
such many-body states, we use unitarity to replace the sum
over these states with the contribution from $M_i$ themselves:
\begin{equation}
\sum {|M_i~{\rm decay~products}\rangle \langle M_i ~{\rm decay~products}|}
=|M_i\rangle \langle M_i|
\end{equation}
As in ref.\cite{Zen1}, we restrict our study to the case
when  FSI
(now Zweig-rule-satisfying) are SU(3)-symmetric.
Consequently, for
the intermediate $M_1M_2$ states it is appropriate 
to use the basis
in which a pair of mesons $M_1M_2$ (in most cases, heavy)
 forms a state belonging 
to a definite SU(3) multiplet. In ref.\cite{Zen1} 
the amplitudes for the short-distance-driven decay of a $B$ meson into
such states
are expressed in terms of standard
SD diagram amplitudes $T,T'$ (tree), $C,C'$ (color
suppressed), $P,P'$ (penguin), $E,E'$ (exchange), $A,A'$ (annihilation),
$PA,PA'$ (penguin annihilation), $S,S'$ (singlet penguin), 
$SS,SS'$ (double singlet penguin). As usual, strangeness-conserving
$\Delta S =0$ (strangeness-violating $|\Delta S| =1$)
processes are denoted by unprimed (primed) amplitudes.
These amplitudes may be thought to incorporate the electroweak penguin
contributions according to $T \to T+P^c_{EW}$, $P \to P-P^c_{EW}/3$,
$C \to C+P_{EW}$, $S \to S-P_{EW}/3$ \cite{elpenguins}.

Let us first recapitulate
the essential assumption of ref.\cite{Zen1}, which permits parametrization
of all FSI effects in terms of a few parameters only.
Namely, since at the SD level it is not yet decided whether the
particular quark-level state will hadronize as the $PP$ state or one
of the heavier $M_1M_2$ states,
 one expects that quark-level SD amplitudes for 
$B$ decays
into two arbitrary mesons $M_1M_2$
are proportional to the corresponding amplitudes of SD decay into two 
pseudoscalar mesons $P_1P_2$.
The coefficient of proportionality may
depend on the type of mesons in the $M_1M_2$ state
(ie. whether $M_1$ ($M_2$) 
are vector, axial, tensor, etc.), but 
- by virtue of the SU(3) symmetry - it must be the same
for all $M_1$, $M_2$ within given SU(3) multiplets. Consequently,
this coefficient
 may be absorbed into the FSI amplitude for the process $M_1M_2 \to P_1P_2$,
  whose size, due to our ignorance, must be again
  treated as a free parameter.
In other words, $T,C,P,..$-type amplitudes 
used for SD $B \to M_1M_2$ decays 
are normalized to their SD $B \to P_1P_2$ counterparts. This justifies
the use of the same
letter $T$  for both $T_{B \to P_1P_2}\equiv T$ and
$T_{B \to M_1M_2}$, and similarly for other
types of diagrams. 
(A part of the 
analysis of this paper would go through also if coefficients of
proportionality between the $B \to M_1M_2$ and $B \to P_1P_2$ amplitudes
depended on the type of diagram, ie. if they were different for tree, penguin,
etc. amplitudes. Since this introduces additional parameters, we
do not consider this possibility further on.)
When the
 rescattering contributions from all
intermediate $M_1M_2$ states are added, they are gathered into
a few groups differing in their
SU(3) symmetry structure. 
For each such group, the SU(3) structure is factorized
and then the remaining sum of unknown free parameters is replaced
with a single parameter, as discussed in \cite{Zen1}.

The amplitudes for $B \to M_1M_2$, calculated in \cite{Zen1},  are
given here in Tables \ref{table27}, \ref{table8} and \ref{table1}
(in a normalization adjusted to that used normally for $B \to P_1P_2$).
From these Tables the amplitude of, say, a $B^+$ decay into a pair
of two mesons $M_1M_2$  in an overall 
antisymmetric octet state and of total isospin $1/2$ may be read of
from the column marked $(8_a,1/2)$ to be
 $\frac{1}{\sqrt{3}}(T'-C'+3P'+3A')$.

\begin{table}[p]
\caption{SD amplitudes into two-meson 
octet-octet states forming a 27-plet}
\label{table27}
\begin{center}
\begin{footnotesize}
\begin{tabular}{cccccc}
\hline
 &(27,2) & (27,3/2) & (27,1) & (27,1/2) & (27,0) \\
\hline
$B^+$   &$-\frac{1}{\sqrt{2}}(T+C)$ & $\frac{1}{\sqrt{3}}(T'+C')$ 
& $-\frac{1}{\sqrt{10}}(T+C)$ & $\frac{2}{\sqrt{15}}(T'+C')$ &
$0$ \\
$B^0_d$ & $-\frac{1}{\sqrt{3}}(T+C)$ & $\frac{1}{\sqrt{3}}(T'+C')$ &
$0$ & $\frac{1}{\sqrt{15}}(T'+C')$ & $-\frac{1}{2\sqrt{15}}(T+C)$ \\
$B^0_s$ & $0$ & $-\frac{1}{\sqrt{3}}(T+C)$ & $-\frac{1}{\sqrt{5}}(T'+C')$ & 
$-\frac{1}{\sqrt{15}}(T+C)$ & $\sqrt{\frac{3}{20}}(T'+C')$ \\
\hline
\end{tabular}
\end{footnotesize}
\end{center}
\end{table}

\begin{table}[p]
\caption{SD amplitudes into two-meson states in overall octet SU(3)
representation for symmetric octet-octet,
antisymmetric octet-octet, and (symmetric) octet-singlet combinations}
\label{table8}
\begin{center}
\begin{footnotesize}
\begin{tabular}{cccc}
\hline
 & $(8_s,1)$ & $(8_s,1/2)$ & $(8_s,0)$\\
\hline
$B^+$    &$-\frac{1}{\sqrt{15}}(T+C+5 P +5 A)$ & 
$\frac{1}{\sqrt{15}}(T'+C'+5 P'+5 A')$ &  $0$\\
$B^0_d$  & $\sqrt{\frac{5}{6}}(E-P)$ & $\frac{1}{\sqrt{15}}(3T'-2C'+5P')$ &
$-\frac{1}{3\sqrt{10}}(6T-4C+5P+5E)$ \\
$B^0_s$  & $\frac{1}{\sqrt{30}}(3T'+5E'-2C')$ & 
$-\frac{1}{\sqrt{15}}(3T-2C+5P)$ & $\frac{1}{3\sqrt{10}}(3T'-2C'+10P'-5E')$ \\
\hline
\hline
& $(8_a,1)$ & $(8_a,1/2)$ & $(8_a,0))$ \\
\hline
$B^+$   & $-\frac{1}{\sqrt{3}}(T-C+3 P+3 A)$ & 
$\frac{1}{\sqrt{3}}(T'-C'+3 P'+3 A')$ & $0$ \\
$B^0_d$ & $-\frac{1}{\sqrt{6}}(2T+3P-3E)$ & 
$\frac{1}{\sqrt{3}}(T'+3P')$ & $-\frac{1}{\sqrt{2}}(E+P)$ \\
$B^0_s$ & $-\frac{1}{\sqrt{6}}(T'-3E')$ & 
$-\frac{1}{\sqrt{3}}(T+3P)$ & $\frac{1}{\sqrt{2}}(T'+2P'-E')$ \\
\hline
\hline
& $(8_{\{81\}},1)$ & $(8_{\{81\}},1/2)$ & $(8_{\{81\}},0)$ \\
\hline
$B^+$   & $-\frac{1}{\sqrt{3}}(T+C+2P+2A+3S)$ &
$\frac{1}{\sqrt{3}}(T'+C'+2P'+2A'+3S') $ & $ 0$\\
$B^0_d$ & $\frac{1}{\sqrt{6}}(-2P+2E-3S)$ & 
$\frac{1}{\sqrt{3}}(C'+2P'+3S')$ & 
$-\frac{1}{3\sqrt{2}}(2C+2P+2E+3S)$\\
$B^0_s$ & $\frac{1}{\sqrt{6}}(C'+2E')$ & 
$-\frac{1}{\sqrt{3}}(C+2P+3S)$ & 
$-\frac{1}{3\sqrt{2}}(-C'-4P'+2E'-6S')$\\
\hline
\end{tabular}
\end{footnotesize}
\end{center}
\end{table}

\begin{table}[p]
\caption{SD amplitudes into two-meson octet-octet and singlet-singlet
states forming an overall singlet}
\label{table1}
\begin{center}
\begin{footnotesize}
\begin{tabular}{ccc}
\hline
& $(1_{88},0)$ & $(1_{11},0)$ \\
\hline
$B^+$    & $0$ & $0$ \\
$B^0_d$  & $\frac{1}{6}(3T-C+8P+8E+12 PA)$ & 
$\frac{1}{3\sqrt{2}}(2C+2P+2E+3 PA+6S +SS)$ \\
$B^0_s$  & $\frac{1}{6}(3T'-C'+8P'+8E'+12 PA')$ & 
$\frac{1}{3\sqrt{2}}(2C'+2P'+2E'+3PA'+6S'+SS')$ \\
\hline
\end{tabular}
\end{footnotesize}
\end{center}
\end{table}

\section{General FSI amplitudes satisfying Zweig rule}

Zweig-rule-satisfying rescattering $M_1M_2 \to P_1P_2$ 
is described by two types of connected
diagrams:
the "uncrossed" diagrams of Fig. 1($u$), and the "crossed" diagrams of 
Fig. 1($c$).
By virtue of Bose statistics, the final $P_1P_2$ pair must be in 
an overall symmetric state.

For the uncrossed $M_1M_2 \to P_1P_2$
diagrams, the requirement of Bose statistics for $P_1P_2$
means that there are two allowed
types of 
SU(3) amplitudes, ie. 
(using a particle symbol for the
corresponding SU(3) matrix):
\begin{equation}
\label{uncrosseds}
{\rm Tr}(\{M_1^{\dagger},M_2^{\dagger}\}\{P_1,P_2\})~u_+
\end{equation} 
and
\begin{equation}
\label{uncrosseda}
{\rm Tr}([M_1^{\dagger},M_2^{\dagger}]\{P_1,P_2\})~u_-
\end{equation}
where the requirement in question
 is reflected through the presence of the anticommutator
$\{P_1,P_2\}$ of meson matrices,
and $u_{\pm}$ denote the strength of rescattering amplitudes.
Eqs.(\ref{uncrosseds},\ref{uncrosseda}) incorporate
nonet symmetry for both intermediate and final mesons.
Invariance of  strong
interactions under charge conjugation demands that
mesons $M_1$ and $M_2$ belong to multiplets of the same (opposite)
C-parities for
the first (second) amplitude above.
Thus, the subscript of $u_{\pm}$ may be understood also as the value of
the product of the C-parities of mesons $M_1$ and $M_2$.

For the crossed diagrams, the requirement of $P_1 \rightleftharpoons P_2$
symmetry admits only one combination:
\begin{equation}
\label{crossed}
{\rm Tr}(M_1^{\dagger}P_1M_2^{\dagger}P_2
+M_1^{\dagger}P_2M_2^{\dagger}P_1)~c
\end{equation}
where $c$ denotes the strength of the amplitude. 
This combination, symmetric under $M_1\rightleftharpoons M_2$,
is charge-conjugation invariant if $M_1$ and $M_2$ have 
C-parities of the same sign.
(If $M_1$ and $M_2$ have opposite C-parities, 
charge-conjugation-invariance requires that the "$+$" sign in
Eq.(\ref{crossed}) be changed into "$-$".  This leads to an expression
antisymmetric under $P_1 \rightleftharpoons P_2$, in violation
of the requirement of Bose statistics.)

\section{Rescattering contributions to B decays}
From Eqs.(\ref{uncrosseds},\ref{uncrosseda},\ref{crossed})
one can evaluate the contribution of $u$-type
FSI diagrams with intermediate two-meson
states in each of the different SU(3) representations: $27$, $8_s$, $8_a$,
$8_{\{81\}}$, $1_{\{88\}}$, and $1_{\{11\}}$. 
The set
consisting of $8_s$, $8_{\{81\}}$, $1_{\{88\}}$, 
and $1_{\{11\}}$
originates from the intermediate
states in which mesons $M_1$ and $M_2$ have the same C-parity, while
the other set (ie. $8_a$) - in which they have opposite C-parities.
Their respective sizes are measured by  $u_+$ and $u_-$.
 In the following we will use
 \begin{eqnarray}
 u & \equiv & (u_++u_-)/2\\
 d &\equiv & u_+-u_-
 \end{eqnarray}
 It was argued that
it is the {\em sum} over many intermediate states that
might lead to significant FSI effects. This is represented by $u$.
The terms proportional to $d$ represent the {\em difference} of
contributions from the $C_1C_2=+1$ and $C_1C_2=-1$ states.
While such difference for
 the lowest-lying $P_1P_2$ ($C_1C_2=+1$) 
and $P_1V_2$ ($C_1C_2=-1$, $V$ - vector meson) states
may be important in itself, these are
just two of many possible intermediate states.
Thus, hopefully, the contribution of the lowest states to the difference in
question
is not large.  
For heavier $M_1M_2$ states, one may expect that the
difference between the contributions from many neighbouring and overlapping
$C_1C_2=+1$ and $C_1C_2=-1$ intermediate states
is small. Thus, unless the few lowest-lying intermediate states
strongly violate the expected approximate equality of the
$C_1C_2=+1$ and $C_1C_2=-1$ contributions,  
$|d| $ should be much smaller than $|u|$.

 Summation over all
 SU(3) representations in the $s$-channel
yields expressions for the combined contributions induced by
all uncrossed and all crossed FSI diagrams.
For the decays $B \to \pi \pi, \pi K, K \bar{K}$, these 
contributions are
given in Tables \ref{tableDeltaS0} and \ref{tableDeltaS1}.
Relationship between $u_+$, $u_-$, $c$, and the 
parameters defined in ref.\cite{Zen1} is as follows:
\begin{eqnarray}
f_{27}&=&2c\\
f_s&=&\frac{5}{3}u_+-\frac{4}{3}c\\
f_a     &=&-5u_-\\
f_{88}  &=&\frac{16}{3}u_+-\frac{2}{3}c\\
f_{18}  &=&\frac{10}{3}u_++\frac{10}3{}c\\
f_{11}  &=&\frac{4}{3}u_++\frac{4}{3}c
\end{eqnarray}  
or, equivalently,
\begin{eqnarray}
\Delta _1 &=& 5d \\
\Delta _2 &=& 15d -10 c\\
\Delta _3 &=& 10 u \\
\Delta _4 &=& 20 u + 10 d -10 c\\
\Delta _5 &=& 60 u + 30 d - 20 c
\end{eqnarray} 

\begin{table}[t]
\caption{Contributions to $\Delta S =0$ decays 
$B \to \pi \pi, \pi K, K\bar{K}$ }
\label{tableDeltaS0}
\begin{center}
\begin{footnotesize}
\begin{tabular}{cccc}
\hline
decay & SD  & u-type FSI diagrams & c-type FSI diagrams\\
\hline
$B^+ \to \pi ^+ \pi ^0  $   & $-\frac{1}{\sqrt{2}}(T+C)$ &
$0$ & $-\frac{1}{\sqrt{2}}(T+C)\cdot 2c$\\
$B^+ \to K ^+ \bar{K}^0 $   & $-P$ & 
$-(C\cdot 2u + (T+3P)d )$ & $0$ \\
\hline 
$B^0_d \to \pi ^+ \pi ^-  $ & $-(T+P)$ & 
$-((T+2P)\cdot 2u + (T+3P)d)$ & 
$-C \cdot 2c$ \\
$B^0_d \to \pi ^0 \pi ^0$ & $-\frac{1}{\sqrt{2}}(C-P)$ & 
$\frac{1}{\sqrt{2}}((T+2P)\cdot 2u+(T+3P)d )$ & 
$-\frac{1}{\sqrt{2}}T\cdot 2c$  \\
$B^0_d \to K^+ K^-$ & $0$ & $(T+2P) \cdot 2u $ & $0$ \\
$B^0_d \to K^0 \bar{K}^0$ & $-P$ & $-(2P \cdot 2u +(T+3P)d )$ & $0$\\
\hline
$B^0_s \to \pi ^+ K^-$ & $-(T+P)$ & $-(T+3P)d $ & $-C\cdot 2c$ \\
$B^0_s \to \pi ^0 \bar{K}^0$ & $-\frac{1}{\sqrt{2}}(C-P)$ &
$\frac{1}{\sqrt{2}}(T+3P)d $ & $-\frac{1}{\sqrt{2}}T\cdot 2c $ \\
\hline
\end{tabular}
\end{footnotesize}
\end{center}
\end{table}

\begin{table}[t]
\caption{Contributions to $|\Delta S =1|$ 
decays $B \to \pi \pi, \pi K, K\bar{K}$}
\label{tableDeltaS1}
\begin{center}
\begin{footnotesize}
\begin{tabular}{cccc}
\hline
decay & SD  & u-type FSI diagrams & c-type FSI diagrams\\
\hline
$B^+ \to \pi ^+ K ^0  $   & $-P'$ &
$-((T'+3P'+S')d +(C'+S')\cdot 2u)$ & 
$-S'\cdot 2c$\\
$B^+ \to \pi ^0 K^+ $   & $\frac{1}{\sqrt{2}}(T'+C'+P')$ & 
$\frac{1}{\sqrt{2}}((T'+3P'+S')d +(C'+S')\cdot 2u)$ & 
$\frac{1}{\sqrt{2}}(T'+C'+S')\cdot 2c$ \\
\hline 
$B^0_d \to \pi ^- K^+  $ & $T'+P'$ & 
$(T'+3P'+S')d + S'\cdot 2u$ & $(C'+S')\cdot 2c$ \\
$B^0_d \to \pi ^0 K^0$ & $\frac{1}{\sqrt{2}}(C'-P')$ & 
$-\frac{1}{\sqrt{2}}((T'+3P'+S')d +S'\cdot 2u)$ & 
$\frac{1}{\sqrt{2}}(T'-S')\cdot 2c$  \\
\hline
$B^0_s \to \pi ^+ \pi ^-$ & $0$ & $-(T'+2P') \cdot 2u $ & $0$ \\
$B^0_s \to \pi ^0 \pi ^0$ & $0$ & $\frac{1}{\sqrt{2}}(T'+2P')\cdot 2u $ 
& $0$\\
$B^0_s \to K^+ K^-$ & $T'+P'$ & $(T'+2P'+S')\cdot 2u+(T'+3P'+S')d  $ & 
$(C'+S')\cdot 2c$ \\
$B^0_s \to K^0 \bar{K}^0$ & $-P'$ &
$-((2P' +S')\cdot 2u +(T'+3P'+S')d )$ & 
$-S'\cdot 2c$ \\
\hline
\end{tabular}
\end{footnotesize}
\end{center}
\end{table}

In the $\Delta S =0$ decays, 
we keep only the terms proportional to $T$, $C$, and $P$, since these
are expected to be
the leading ones. 
In the $|\Delta S|=1 $ decays, the contributions from the singlet
penguin $S'$ are shown as well.

For completeness, it is appropriate to discuss the decays
into $\pi \eta , \pi \eta ', K \eta $, and $K \eta '$ as well. 
FSI-induced contributions to these decays are gathered
in Tables
\ref{tableDeltaS0eta} and
\ref{tableDeltaS1eta}.
For $\eta $ and $\eta '$ we used
\begin{eqnarray}
\eta   &=& \frac{1}{\sqrt{3}}(u\bar{u}+d\bar{d}-s\bar{s})\\
\eta ' &=& \frac{1}{\sqrt{6}} (u\bar{u}+d\bar{d}+2s\bar{s})
\end{eqnarray}
corresponding to an octet-singlet mixing angle of $\theta = -19.5^o$ 
(as generally assumed, see eg.
  \cite{Lipkin81,Chau91D43,GR96D53,DGR97PRL79}).
When calculating expressions in Table \ref{tableDeltaS0eta},
as in Table \ref{tableDeltaS0}
 we have assumed that the
SD singlet penguin amplitude is negligible.  
The large branching ratios of $B \to K\eta '$ decays 
seem to require significant
$S'$.
Thus, we have kept $S'$ in Table \ref{tableDeltaS1eta} not only in the SD
contribution but also, as in Table \ref{tableDeltaS1}, in the FSI part.

In principle, 
if FSI are important, the decay $B \to K \eta '$ may be described
also when the short-distance singlet
penguin $S'$ is small. 
Indeed, 
neglecting all terms in the FSI contributions but those proportional to $P' $,
and putting $d = 0$ for simplicity,
we observe that an effective singlet penguin amplitude
of size $S'_{eff} \approx P'\cdot 2c $ is generated
 in the formulas for $B^+,B^0_d \to K\eta, K\eta '$
amplitudes
 (with the $P'$ amplitude defined
from $B^+ \to \pi ^+K^0$
and $B^0_d \to \pi ^- K^+$ decays, see Table \ref{tableDeltaS1}).
Thus, a part of the singlet penguin
amplitude, phenomenologically required in $B^+, B^0_d \to K \eta '$ decays, 
may have its origin in final state interactions.
A positive value of $c$ around $+0.24$ brings about a constructive
interference between $P'$ and $S'_{eff}$, 
and permits a fit to the data \cite{DGR97PRL79,CR2001}.
In fact, if a large part of the effective singlet penguin amplitude
 originates from FSI, one expects its phase to be real with respect to that
of the regular penguin: 
the $qq\bar{q}\bar{q}$ 
structure of $s$-channel states in the crossed diagrams 
should entail real $c$.

\begin{table}[t]
\caption{Contributions to $\Delta S =0$ decays 
$B^+, B^0_d \to \pi \eta , \pi \eta '$ and 
$B^0_s \to \bar{K} \eta, \bar{K} \eta '$}
\label{tableDeltaS0eta}
\begin{center}
\begin{footnotesize}
\begin{tabular}{cccc}
&&&\\
\hline
decay & SD  & u-type FSI diagrams & c-type FSI diagrams\\
\hline
$B^+ \to \pi ^+ \eta  $   & $-\frac{1}{\sqrt{3}}(T+C+2P)$ &
$-\frac{2}{\sqrt{3}}(C\cdot 2u +(T+3P)d )$ & 
$-\frac{1}{\sqrt{3}}(T+C+P)\cdot 2c$\\
$B^+ \to \pi ^+ \eta '$   & $-\frac{1}{\sqrt{6}}(T+C+2P)$ & 
$-\frac{2}{\sqrt{6}}(C\cdot 2u + (T+3P)d )$ & 
$-\frac{1}{\sqrt{6}}(T+C+4P)\cdot 2c$ \\
\hline 
$B^0_d \to \pi ^0 \eta  $ & $-\frac{2}{\sqrt{6}}P$ & 
$\frac{2}{\sqrt{6}}(T\cdot 2u - (T+3P)d)$ & 
$-\frac{1}{\sqrt{6}}P\cdot 2c$ \\
$B^0_d \to \pi ^0 \eta '$ & $-\frac{1}{\sqrt{3}}P$ & 
$\frac{1}{\sqrt{3}}(T\cdot 2u-(T+3P)d )$ & 
$-\frac{2}{\sqrt{3}}P\cdot 2c$  \\
\hline
$B^0_s \to \bar{K}^0 \eta$ & $-\frac{1}{\sqrt{3}}C$ & $0$ & 
$-\frac{1}{\sqrt{3}}(T+P)\cdot 2c$\\
$B^0_s \to \bar{K}^0 \eta '$ & $-\frac{1}{\sqrt{6}}(C+3P)$ & 
$-\frac{1}{\sqrt{6}}(T+3P)\cdot 3 d $ & 
$-\frac{1}{\sqrt{6}}(T+4P)\cdot 2c $ \\
\hline
\end{tabular}
\end{footnotesize}
\end{center}
\end{table}

\begin{table}[t]
\caption{Contributions to $|\Delta S =1|$ decays 
$B^+, B^0_d \to K \eta , K \eta ' $ and 
$B^0_s \to \pi \eta , \pi \eta '$}
\label{tableDeltaS1eta}
\begin{center}
\begin{footnotesize}
\begin{tabular}{cccc}
&&&\\
\hline
decay & SD  & u-type FSI diagrams  & c-type FSI diagrams\\
\hline
$B^+ \to K^+ \eta$   & $\frac{1}{\sqrt{3}}(T'+C'+S')$ &
$0$ &
$\frac{1}{\sqrt{3}}(T'+C'+P')\cdot 2c$ \\
$B^+ \to  K^+ \eta '$ & $\frac{1}{\sqrt{6}}(T'+C'+3P'+4S')$ &
$\frac{3}{\sqrt{6}}((C'+S')\cdot 2u+(T'+3P'+S') \delta)$ &
$\frac{1}{\sqrt{6}}(T'+C'+4P'+3S')\cdot 2c$ \\
\hline
$B^0_d \to  K^0 \eta$ & $\frac{1}{\sqrt{3}}(C'+S')$ &
$0$ & $\frac{1}{\sqrt{3}}(T'+P')\cdot 2c $ \\
$B^0_d \to  K^0 \eta '$ & $\frac{1}{\sqrt{6}}(C'+3P'+4S')$ &
$\frac{3}{\sqrt{6}}(S'\cdot 2u+(T'+3P'+S')\delta )$ & 
$\frac{1}{\sqrt{6}}(T'+4P'+3S')\cdot 2c $\\
\hline
$B^0_s \to \pi ^0 \eta $ & $-\frac{1}{\sqrt{6}} C'$ & 
$\frac{2}{\sqrt{6}}T'\cdot 2u$ & $-\frac{1}{\sqrt{6}}T'\cdot 2c $ \\
$B^0_s \to \pi ^0 \eta '$ & $\frac{1}{\sqrt{3}}C' $ & 
$\frac{1}{\sqrt{3}}T'\cdot 2u $ & $\frac{1}{\sqrt{3}} T' \cdot 2c$ \\
\hline
\end{tabular}
\end{footnotesize}
\end{center}
\end{table}

Let us now analyse the formulas given in Tables
\ref{tableDeltaS0}-\ref{tableDeltaS1eta} 
 to see what general FSI pattern is generated, and
 whether something can be said not only about $c$ but also
about the values of $u$, and $d$,
 and the extraction of the SD amplitudes
from the data. To this end, in Tables 
\ref{tableDeltaS0ren} and \ref{tableDeltaS1ren}
we rewrite the content of Tables
\ref{tableDeltaS0}-\ref{tableDeltaS1eta} in terms 
of redefined amplitudes
\begin{eqnarray}
\label{tilderenbeg}
\tilde{T}' & = & T'+ C'\cdot 2c\\
\tilde{C}' & = & C'+ T'\cdot 2c\\
\label{primePnew}
\tilde{P}' & = & P' +S'\cdot (2c+2u)+(T'+3P'+S')d \\
\label{tilderenend}
\tilde{S}' & = & S'+ P'\cdot 2c
\end{eqnarray}
and (for $S=0$)
\begin{eqnarray}
\label{tilderenbegunpr}
\tilde{T} & = & T+ C\cdot 2c\\
\tilde{C} & = & C+ T\cdot 2c\\
\label{Pnew}
\tilde{P} & = & P +(T+3P)d\\
\label{tilderenendunpr}
\tilde{S} & = & P\cdot 2c.
\end{eqnarray}

Note that for $d\ne 0$ the redefined amplitude $\tilde{P}$ 
depends on two weak phases: $\beta$ and $\gamma$.
This may affect the methods of $\gamma $ determination (see below).
The analysis of Tables \ref{tableDeltaS0ren} and \ref{tableDeltaS1ren},
and of the general short-distance expressions for 
decay amplitudes (eg. ref.\cite{GHLR94,Zen1}) shows that
all  inelastic FSI effects marked in the Tables as "observable FSI
modifications"
have the pattern of
 effective annihilation $\tilde{A}$, exchange $\tilde{E}$,
and penguin annihilation $\tilde{PA}$
amplitudes
\begin{eqnarray}
\tilde{A}&=&C\cdot 2u\\
\tilde{E}&=&T\cdot 2u\\
\tilde{PA}&=&2P\cdot 2u
\end{eqnarray}
and
\begin{eqnarray}
\tilde{A}'&=&C'\cdot 2u\\
\tilde{E}'&=&T'\cdot 2u\\
\tilde{PA}'&=&2P'\cdot 2u
\end{eqnarray}
Thus, for SU(3)-symmetric FSI 
it is impossible to distinguish between pure SD 
and FSI-corrected amplitudes
on the basis of experimental data alone.
Only if  the short-distance $A,E,...$
amplitudes are known to be negligible, may one attempt to deduce
the size of FSI effects.
In that case, $u$ might be estimated as follows.
From $B^+ \to \pi ^+ K^0$, neglecting the $C'\cdot 2u$ term,
 one extracts $\tilde{P}' \approx 4.15$
 (with the amplitude squares giving $B$ decay 
branching ratios in units of $10^{-6}$).
If terms proportional to $d $ and $S'$ in 
Eqs(\ref{primePnew},\ref{Pnew}) may be neglected, one finds
$|\tilde{P}| \approx |V_{td}/V_{ts}| |\tilde{P}'| \approx 
0.179 |\tilde{P}'| = 0.74 $ with an error of $\pm 0.06$.
Neglecting $\tilde{C}$ with respect to $\tilde{T}$ 
from the $B^+ \to \pi ^+ \pi ^0$ decays one obtains that
$|\tilde{T}|$ is $ 3.38 $.
Various analyses tend to give a slightly smaller central value:
$|\tilde{T}|=2.7\pm0.6$
\cite{DGR97PRL79,CR2001}, which is used below.
It follows that $|\tilde{T}'| \approx 
|V_{us}/V_{ud}||f_K/f_{\pi }| |\tilde{T}| \approx 0.74 \pm 0.16$.
Since $|T'|\ll |P'|$ one may neglect $T'$
in the amplitude describing $B^0_s \to \pi ^+ \pi ^-$
decays  to find ($\Gamma $ denoting the branching ratio)
\begin{equation}
\label{usize}
|u|^2=\frac{1}{4}\frac{\Gamma (B^0_s \to \pi ^+ \pi ^-)}
{\Gamma (B^+ \to \pi ^+ K^0)}
\end{equation}
Then, from
\begin{equation}
\frac{\Gamma (B^0_s \to K^0 \bar{K}^0)}{\Gamma (B^+ \to \pi ^+ K^0)}=|1+4u|^2
=\frac{\Gamma (B^0_d \to K^0 \bar{K}^0)}{|P|^2}
\end{equation}
one might deduce the phase $\delta _u$ of $u$.
Finally, the size of $\Gamma (B^0_d \to K^+ K^-)$,
with $|u|$ known and
$|\tilde{T}|$ being comparable to $2|\tilde{P}|$,
puts a constraint on the relative phase of $
\tilde{T}$ and $ \tilde{P}$. 

In the absence of $B^0_s$ decay data, the best that can be done is to
place an upper limit on the size of $|u|$, eg.
 by measuring the branching ratios for the
$B^0_d \to K^+K^- (K^0 \bar{K}^0)$ decays \cite{GR98} .
The estimate of $|u|$ from 
the correction term $(T+2P)\cdot 2u$
in $B^0_d \to K^+K^-$ can be hampered if $T$ and $P$ interfere
destructively, as might be the case \cite{GR0109238}.
In Fig. 2 we show the present bounds on the size of $|u|$ and the SD
tree-penguin relative phase $\phi _T -\phi _P$, 
obtained from the $B^0_d \to \pi^+\pi^-$ entry in Table 8 
when the central value of $4.4$ is assumed for
 the corresponding branching ratio.
The dependence of rescattering parameter $|u|$ on the tree-penguin relative
phase $\phi _t-\phi _p$ is given there for several cases:\\
a) for central values $|\tilde{P}|=0.74$ and $|\tilde{T}|=2.7$ -
thick solid line for $\cos \delta _u = -1$, thick dashed line for
$\cos \delta _u = -0.85$,\\ 
b)  
for $|\tilde{P}|_{max}=0.80$, 
$|\tilde{T}|_{min}=2.10$, and $\cos \delta _u = -1$ -
short dashed line (approximate lower range),\\
c) 
for $|\tilde{P}|_{min}=0.68$, 
$|\tilde{T}|_{max}=3.30$, and $\cos \delta _u =-0.85$ -
long dashed line (approximate upper range), \\
d) the upper bound on $|u|$ from $B \to K^+K^-$ branching ratio 
(using central values for $|\tilde{T}|$ and $|\tilde{P}|$)-
thin line: $B.R.(B \to K^+K^-)=1.9\cdot 10^{-6}$ (present experimental value);
dotted line: $B.R.(B \to K^+K^-)=0.1\cdot 10^{-6}$.

Fig. 2 shows that $|u|$ of the order of $0.15$ is consistent with the present
data for any value of $|\phi _T-\phi _P|$. 
If $|u|$ is close to $0$,  the data seem to indicate a
cancellation between the tree and
penguin diagrams.
(Although Fig. 2 shows that 
for $|\tilde{T}|=2.7$, $|\tilde{P}|=0.74$
and the $B^0_d \to \pi^+\pi^-$ branching ratio of $4.4$
there is 
no solution for $|\phi _T-\phi _P|$ above $150^o$;
such a solution does exist if errors on $\tilde{T}$, etc. are admitted.)

\section{FSI-induced errors in the extraction of $\gamma $}
Let us now comment on the relationship
between the FSI effects in $B^+ \to K^+\bar{K}^0$,
$B^+ \to \pi ^+ K^0$, $B^0_d \to \pi ^- K^+$, and 
$B^0_s \to \pi ^+ K^-$ discussed in ref.\cite{Zen1}.
It was argued there that the FSI effects may affect
the determination of the CP-violating angle $\gamma$ from the 
latter three decays.
These conclusions followed from the neglect of FSI-induced terms
originating from SD-driven diagrams other than $P$, $P'$, $T$, and 
$T'$, and from the subsequent discussion of the four
relevant amplitudes in the case when FSI corrections proportional
to $T$ are kept, while those proportional to $T'$ are neglected.
When Zweig rule is maintained in FSI,
the relevant formulas may be read from
Tables \ref{tableDeltaS0}, \ref{tableDeltaS1}
to be
\begin{eqnarray}
\label{FSIgamma1}
W(B^+ \to K^+\bar{K}^0)&=&-\bar{P}-Td\\
\label{FSIgamma2}
W(B^0_s \to \pi^+K^-)  &=&-\bar{P}-\bar{T}+2Td\\
W(B^+\to \pi ^+K^0)    &=&-\bar{P}'\\
\label{FSIgamma4}
W(B^0_d \to \pi ^- K^+ ) &=&\bar{P}'+\bar{T}'
\end{eqnarray}
where $\bar{T^{(')}}=T^{(')}(1+3d)$, 
$\bar{P^{(')}}=P^{(')}(1+3d)$.
We observe that the FSI effects discussed in \cite{Zen1} are
proportional to $d $ representing the difference between the overall
 contributions
from the $C_1C_2=+1$ and $C_1C_2=-1$ states.
Since, as discussed earlier,
$d $ is probably much smaller than $u$,
the FSI effects referred to in ref\cite{Zen1} should not
affect the determination of $\gamma $ too much. 

In order to get a theoretical
feeling for what might be the absolute size of $d$, 
let us consider the following.
As discussed earlier, a nonzero value of $d$ arises 
most probably from the few lowest-lying intermediate states.
Consequently, $d$ should be (very roughly) of the order
of a contribution from a single intermediate 
channel, the best representative being the $PP$ state composed
of two pseudoscalar mesons. 
A rough estimate of this contribution was made in ref.\cite{Zen2001}
in the framework of a Regge exchange model.
From 
Eqs.(10) and (16) of ref.\cite{Zen2001}
one may deduce that the contribution 
to the octet Argand amplitude arising
 from the "uncrossed" Regge
exchange ($d$ measures deviation from 
the $C_1C_2=+1 \Leftrightarrow C_1C_2 =-1$ 
symmetry
for {\em uncrossed} diagrams) is
\begin{equation}
a_8({\rm uncrossed}) = -0.030+0.033i
\end{equation}
Since for small rescattering corrections
the FSI-induced modification of the decay amplitude is 
proportional to $S^{1/2}-1\approx ia$
we expect that $|d|$ should be of the order of
\begin{equation}
\label{dfroma}
|d|\approx |a_8| = 0.045
\end{equation}
While the Regge approach may not be the most reliable one, the above
estimate 
can certainly provide an educated guess as to the order of magnitude of $d$.
Note that in this case $|d| \ll 2|c|,2|u|$ if the values of $|c|\approx 0.24$
and some average $|u| $ around 0.1 or 0.15 (cf. Fig. 2) are accepted.

In Fig. 3 we show errors induced by admitting nonzero $|d|$
in the method of 
the determination of $\gamma $ considered in refs.\cite{gamma}.
Solid lines represent the effect discussed by 
Chiang and Wolfenstein
 \cite{CW}, ie. the influence of a nonzero value of CP-violating angle
$\beta $ upon the extracted value of $\gamma $.
When the calculations of ref.\cite{CW} are extended to include the effect of 
the term proportional to $Td$ in Eq. (\ref{FSIgamma2}), one obtains
the following counterparts of Eqs (7) and (8) from \cite{CW}:
\begin{eqnarray}
\label{CW1}
K&=&1+2 \lambda ^2 \frac{\sin \beta \cos \gamma}{\sin (\beta + \gamma)}+
\lambda ^4 \left(  \frac{\sin \beta}{\sin (\beta + \gamma )}
\right)^2\\
\label{CW2}
KR_d&=&1+r^2+2r \cos \delta  \cos \gamma \\
\label{CW3}
KR_s&=&\lambda ^2 +\left( \frac{r}{\lambda } \right)^2-2 r \cos \delta 
\cos \gamma + 2 |d| 
\cos \delta _d\left[  \left( \frac{r}{\lambda}\right) ^2 + 3 \lambda ^2
-4 r \cos \gamma
\right]
\end{eqnarray}
where $\delta = \phi_T-\phi _P$ is the relative phase between
the tree and penguin amplitudes, $r$ is their ratio, and $\lambda = 0.22$.
As in \cite{CW}  phase $\delta$
is set to zero.
The only difference, when compared to ref.\cite{CW}, is the 
presence of the rightmost term in Eq.(\ref{CW3}).
This term leads to
additional errors on $\gamma$. 
In Fig.3 we show the relevant error bands for $|d|=0.01$ and $|d|=0.05$
and $\cos \delta _d =\pm 1$.
Thus,
for $\gamma $ around $50^o$ to $60^o$ the expected FSI-induced errors
are of the order of $\pm 5^o$.

\section{Conclusions}
This paper shows explicitly how and under what assumptions
all inelastic SU(3)-symmetric 
rescattering effects reduce to the redefinition of initial
SD amplitudes, thus permitting the use of a simple
diagram-based description, albeit with certain modifications.
If FSI are important, the phenomenologically extracted diagram
amplitudes do not have to be equal to those of the SD approaches.
If only $T$, $P$, $C$ ($P'$, $T'$, $C'$, $S'$) SD amplitudes are
nonnegligible, SU(3) symmetric rescattering can 
1) generate effective annihilation, exchange and penguin annihilation
amplitudes, 
2) violate the expected relation $\tilde{P}=V_{td}/V_{ts}\tilde{P}'$, 
and
3) complicate the way
in which weak phases are attributed to penguin-like amplitudes.
It is estimated that the FSI-induced error made when extracting weak phase
$\gamma $ from the $B \to K\pi$
amplitudes may be of the order of $\pm 5^o$ for $\gamma $ around
$50^o - 60^o$.

\section{Acknowledgements}
This work was supported in part by the
Polish State Committee for Scientific Research 
grant 5 P03B 050 21.

\begin{table}[p]
\caption{Effective amplitudes for $\Delta S =0$ decays. 
Contributions of standard form (including
some FSI effects) and possibly observable FSI-induced modifications are 
shown separately.}

\label{tableDeltaS0ren}
\begin{tabular}{ccc}
&&\\
\hline
decay & standard form  & observable FSI modifications \\
\hline
$B^+ \to \pi ^+ \pi ^0  $   & $-\frac{1}{\sqrt{2}}(\tilde{T}+\tilde{C})$ &
$0$ \\
$B^+ \to K ^+ \bar{K}^0 $   & $-\tilde{P}$ & 
$-C\cdot 2u $ \\
\hline 
$B^0_d \to \pi ^+ \pi ^-  $ & $-(\tilde{T}+\tilde{P})$ & 
$-(T+2P)\cdot 2u $ \\
$B^0_d \to \pi ^0 \pi ^0$ & $-\frac{1}{\sqrt{2}}(\tilde{C}-\tilde{P})$ & 
$\frac{1}{\sqrt{2}}(T+2P)\cdot 2u$  \\
$B^0_d \to K^+ K^-$ & $0$ & $(T+2P) \cdot 2u $  \\
$B^0_d \to K^0 \bar{K}^0$ & $-\tilde{P}$ & $-2P \cdot 2u $ \\
\hline
$B^0_s \to \pi ^+ K^-$ & $-(\tilde{T}+\tilde{P})$ & $0$  \\
$B^0_s \to \pi ^0 \bar{K}^0$ & $-\frac{1}{\sqrt{2}}(\tilde{C}-\tilde{P})$ &
$0 $ \\
\hline
$B^+ \to \pi ^+ \eta  $   & 
$-\frac{1}{\sqrt{3}}(\tilde{T}+\tilde{C}+2\tilde{P}+\tilde{S})$ &
$-\frac{2}{\sqrt{3}}C\cdot 2u $ \\
$B^+ \to \pi ^+ \eta '$   & 
$-\frac{1}{\sqrt{6}}(\tilde{T}+\tilde{C}+2\tilde{P}+4\tilde{S})$ & 
$-\frac{2}{\sqrt{6}}C\cdot 2u $  \\
\hline 
$B^0_d \to \pi ^0 \eta  $ & 
$-\frac{1}{\sqrt{6}}(2\tilde{P}+\tilde{S})$ & 
$\frac{2}{\sqrt{6}}T\cdot 2u $  \\
$B^0_d \to \pi ^0 \eta '$ & 
$-\frac{1}{\sqrt{3}}(\tilde{P}+2\tilde{S})$ & 
$\frac{1}{\sqrt{3}}T\cdot 2u$   \\
\hline
$B^0_s \to \bar{K}^0 \eta$ & 
$-\frac{1}{\sqrt{3}}(\tilde{C}+\tilde{S})$ & $0$ \\
$B^0_s \to \bar{K}^0 \eta '$ & 
$-\frac{1}{\sqrt{6}}(\tilde{C}+3\tilde{P}+4\tilde{S})$ & 
$0$ \\
\hline
\end{tabular}
\end{table}

\begin{table}[p]
\caption{Effective amplitudes for $|\Delta S =1|$ 
decays.
Contributions of standard form (including
some FSI effects) and possibly observable FSI-induced modifications are 
shown separately.}
\label{tableDeltaS1ren}
\begin{tabular}{ccc}
&&\\
\hline
decay & standard form  & observable FSI modifications \\
\hline
$B^+ \to \pi ^+ K ^0  $   & $-\tilde{P}'$ &
$-C'\cdot 2u$ \\
$B^+ \to \pi ^0 K^+ $   & 
$\frac{1}{\sqrt{2}}(\tilde{T}'+\tilde{C}'+\tilde{P}')$ & 
$\frac{1}{\sqrt{2}}C'\cdot 2u$ \\
\hline 
$B^0_d \to \pi ^- K^+  $ & $\tilde{T}'+\tilde{P}'$ & 
$0$  \\
$B^0_d \to \pi ^0 K^0$ & $\frac{1}{\sqrt{2}}(\tilde{C}'-\tilde{P}')$ & 
$0$  \\
\hline
$B^0_s \to \pi ^+ \pi ^-$ & $0$ & $-(T'+2P') \cdot 2u $  \\
$B^0_s \to \pi ^0 \pi ^0$ & $0$ & 
$\frac{1}{\sqrt{2}}(T'+2P')\cdot 2u $ 
\\
$B^0_s \to K^+ K^-$ & $\tilde{T}'+\tilde{P}'$ & 
$(T'+2P')\cdot 2u$ \\
$B^0_s \to K^0 \bar{K}^0$ & $-\tilde{P}'$ &
$-2P' \cdot 2u $ \\
\hline
$B^+ \to K^+ \eta$   & 
$\frac{1}{\sqrt{3}}(\tilde{T}'+\tilde{C}'+\tilde{S}')$ &
$0$ \\
$B^+ \to  K^+ \eta '$ & 
$\frac{1}{\sqrt{6}}(\tilde{T}'+\tilde{C}'+3\tilde{P}'+4\tilde{S}')$ &
$\frac{3}{\sqrt{6}}C'\cdot 2u$ \\
\hline
$B^0_d \to  K^0 \eta$ & 
$\frac{1}{\sqrt{3}}(\tilde{C}'+\tilde{S}')$ &
$0$ \\
$B^0_d \to  K^0 \eta '$ & 
$\frac{1}{\sqrt{6}}(\tilde{C}'+3\tilde{P}'+4\tilde{S}')$ &
$0$ \\
\hline
$B^0_s \to \pi ^0 \eta $ & $-\frac{1}{\sqrt{6}} \tilde{C}'$ & 
$\frac{2}{\sqrt{6}}T'\cdot 2u$  \\
$B^0_s \to \pi ^0 \eta '$ & $\frac{1}{\sqrt{3}}\tilde{C}' $ & 
$\frac{1}{\sqrt{3}}T'\cdot 2u $  \\
\hline
\end{tabular}
\end{table}

\vfil

\newpage
FIGURE CAPTIONS

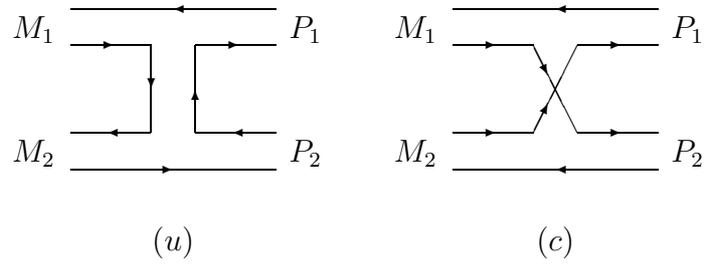
Fig. 1 Types of rescattering diagrams: ($u$) uncrossed, ($c$) crossed

Fig. 2 Dependence of rescattering parameter $|u|$ on tree-penguin relative
phase $|\phi _t-\phi _p|$ 
(a) for central values $|\tilde{P}|=0.74$ and $|\tilde{T}|=2.7$
(thick solid line for $\cos \delta _u = -1$, thick dashed line for
$\cos \delta _u = -0.85$). 
(b) approximate lower range - short-dashed line
(for $|\tilde{P}|_{max}=0.80$, 
$|\tilde{T}|_{min}=2.10$, and $\cos \delta _u = -1$)
(c) approximate upper range - long-dashed line
(for $|\tilde{P}|_{min}=0.68$, 
$|\tilde{T}|_{max}=3.30$, and $\cos \delta _u =-0.85$)
(d) upper bound on $|u|$ from $B \to K^+K^-$ branching ratio -
thin line: $B.R.(B \to K^+K^-)=1.9\cdot 10^{-6}$ (present experimental value);
dotted line: $B.R.(B \to K^+K^-)=0.1\cdot 10^{-6}$.

Fig. 3 Dependence of $\gamma $ on $\beta $. Estimates of FSI-induced errors 
for the method of extracting $\gamma $ from
$B^+, B^0_d, B^0_s \to K\pi$ decays:
(a) $R_d=0.8$, $R_s=0.78$;
(b) $R_d=0.85$, $R_s=0.73$;
(c) $R_d=0.9$, $R_s=0.68$;
(d) $R_d=1.15$, $R_s=0.43$ 

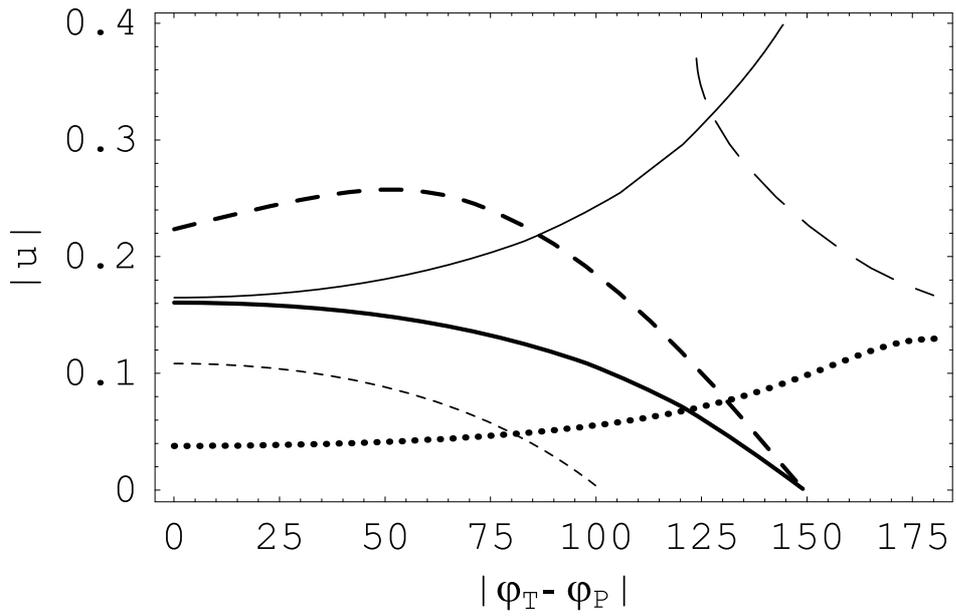
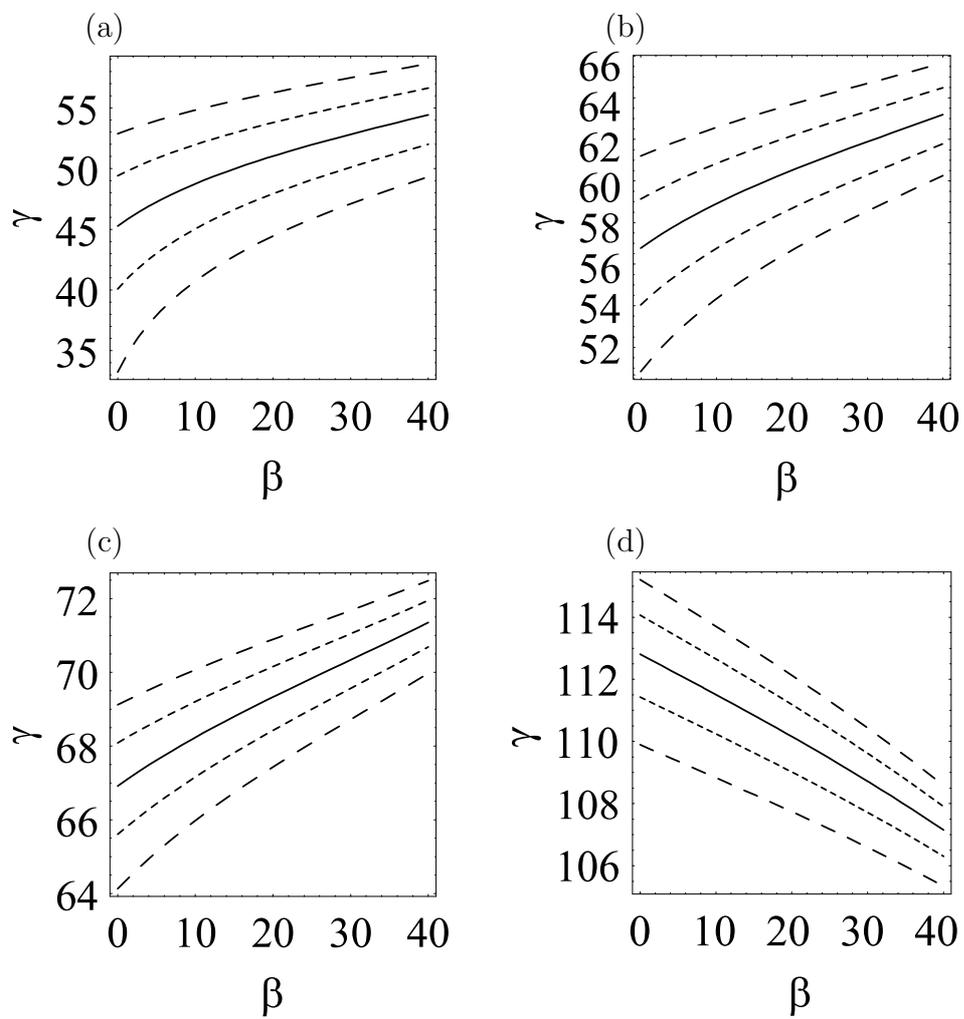
\end{document}